\documentclass{an}
\usepackage{graphicx}
\usepackage{times}
\usepackage{fancyhdr}
\begin{document}

\title{Estimates of the Strouhal number from numerical models of convection \\ \tiny \rm \vspace{0.25cm} $ $Id: paper.tex,v 1.9 2004/10/24 11:06:28 pkapyla Exp $ $}

\author{P.J. K\"apyl\"a\inst{1,2}
\and  M.J. Korpi\inst{3}
\and  M. Ossendrijver\inst{2}
\and  I. Tuominen\inst{1,4}}
\institute{
Astronomy Division, Department of Physical Sciences, P.O. Box 3000, FIN-90014 University of Oulu, Finland
\and 
Kiepenheuer-Institut f\"ur Sonnenphysik, Sch\"oneckstrasse 6, D-79104 Freiburg, Germany
\and 
NORDITA, Blegdamsvej 17, DK-2100, Copenhagen, Denmark
\and 
Observatory, PO BOX 14, FIN-00014 University of Helsinki, Finland}

\date{Received $<$date$>$; 
accepted $<$date$>$;
published online $<$date$>$}

\abstract{ We determine the Strouhal number (hereafter St), which is
  essentially a nondimensional measure of the correlation time, from
  numerical calculations of convection. We use two independent methods
  to estimate St. Firstly, we apply the minimal tau-approximation
  (MTA) on the equation of the time derivative of the Reynolds
  stress. A relaxation time is obtained from which St can be estimated
  by normalising with a typical turnover time. Secondly, we calculate
  the correlation and turnover times separately, the former from the
  autocorrelation of velocity and the latter by following test
  particles embedded in the flow. We find that the Strouhal number is
  in general of the order of 0.1 to 1, i.e. rather large in comparison
  to the typical assumption in the mean-field theories that ${\rm St}
  \ll 1$. However, there is a clear decreasing trend as function of
  the Rayleigh number and increasing rotation. Furthermore, for the
  present range of parameters the decrease of St does not show signs
  of saturation, indicating that in stellar convection zones, where
  the Rayleigh numbers are much larger, the Strouhal number may indeed
  be significantly smaller.   }

\correspondence{petri.kapyla@oulu.fi}

\maketitle

\section{Introduction}
The mean-field theories of hydromagnetic dynamos and angular momentum
transport need knowledge of turbulent correlations, namely the
electromotive force and Reynolds stresses, respectively. The explicit
calculation of these is virtually impossible for any astrophysical
conditions due to the lack of necessary computational cababilities.
Therefore the small-scale effects are usually parametrised by
transport coefficients which relate the turbulent correlations to the
mean quantities (e.g. Steenbeck, Krause \& R\"adler
\cite{SteenKrauRaed1966}; R\"udiger \cite{Ruediger1989}). However, in
order to calculate the transport coefficients, knowledge of the
small-scale velocities and magnetic fields are still needed. Thus,
simplifying assumptions such as the first order smoothing
approximation (hereafter FOSA) have been used (Steenbeck et
al. \cite{SteenKrauRaed1966}). Although the results obtained with FOSA
are in agreement with observations in many cases, the basic
assumptions of the theory have rarely been thoroughly studied
(however, see Petrovay \& Zsarg$\acute{\rm o}$ \cite{PetrZsar1998};
K\"apyl\"a et al. \cite{Kaepylaeea2004b}).

The validity condition for FOSA is that either the Reynolds or the
Strouhal number is small. The former condition is clearly never met,
but of the latter in the context of convection and dynamo theory no
systematic study seems to exist. A further notion associated with the
Strouhal number is that even though if St is not small in the sense
that ${\rm St} \ll 1$, it is still possible to calculate the transport
coefficients from a cumulative series expansion if St $<$ 1
(e.g. Knobloch \cite{Knobloch1978}; Nicklaus \& Stix
\cite{NickStix1988}).

Recently it has been shown numerically that St can \emph{exceed} unity
for forced turbulence (Brandenburg, K\"apyl\"a \& Mohammed
\cite{Brandea04}; Brandenburg \& Subramanian \cite{BrandSubr04}) in
the contexts of passive scalar diffusion and mean-field dynamos,
respectively. In these studies, instead of FOSA, the minimal
tau-approximation (hereafter MTA) is found to be in better agreement
with the calculations. In MTA, instead of the turbulent correlation
itself, the time derivative is investigated, and the higher order
terms are parameterised by a term which is just the original turbulent
correlation divided by a relaxation time (Blackman \& Field
\cite{BlackField2003}). Furthermore, interpreting the relaxation time
as the correlation time of the turbulence the Strouhal number can be
estimated.

The forced turbulence results raise the question of the value of St
for convection, which seems to be rather badly known. For example,
solar surface observations indicate that the lifetime and turnover
time of granules is approximately the same, yielding ${\rm St} \approx
1$ (e.g. Stix \cite{Stix2002}). However, this result may not be
relevant for the solar dynamo which is working in the deeper
layers. In the present study we estimate the Strouhal number for
numerical convection by two independent methods. Firstly we apply the
MTA on the equation of the Reynolds stresses, and secondly we
calculate the correlation and turnover times separately directly from
the flow. The correlation time is estimated from the autocorrelation
of velocity and the turnover time by following test particles embedded
in the flow. The computational model is a Cartesian box situated at a
latitude $\Theta$ on a star. The model is described in detail in
K\"apyl\"a, Korpi \& Tuominen (\cite{Kaepylaeea2004a}).

The remainder of the paper is organised as follows: in
Sect.~(\ref{sec:strouhal}) the two methods used to estimate the
Strouhal number are discussed. Sects.~(\ref{sec:results}) and
(\ref{sec:conclusions}) give the results and conclusions,
respectively.

\section{The Strouhal number}
\label{sec:strouhal}

\subsection{Minimal tau-approximation}
We apply the MTA on the time derivative of the Reynolds stress,
$Q_{ij} = \langle u'_iu'_j \rangle$, where the brackets denote
horizontal averaging and primes the fluctuation. We arrive at the
equation (see K\"apyl\"a et al. \cite{Kaepylaeea2004b} for a more
detailed derivation)
\begin{equation}
  \frac{\partial Q_{ij}}{\partial t} = \Psi_{ijk} \Omega_k - \frac{Q_{ij}}{\tau_{\rm rel}}\;,
  \label{equ:reytau}
\end{equation}
where 
\begin{equation}
  \Psi_{ijk} = -2\,(\epsilon_{ikl} \langle u_j'u_l' \rangle + \epsilon_{jkl} \langle u_i'u_l' \rangle)\;,
\end{equation}
where $\epsilon_{ikl}$ is the Levi-Civita symbol and $\Omega$ the
rotation vector. Taking into account only the lowest order effect for
the vertical $\Lambda$-effect, which is proportional to the component
$Q_{\rm yz}$ in the local convection model, the relaxation time turns
out to be
\begin{eqnarray}
  \tau_{\rm rel} = \frac{Q_{yz}}{2\,\Omega \cos \Theta (Q_{zz} - Q_{yy})}\;. \label{equ:taurel}
\end{eqnarray}
Interpreting $\tau_{\rm rel}$ as the correlation time of the
turbulence, the Strouhal number can be calculated identically as in
the forced turbulence studies (Brandenburg et al. \cite{Brandea04};
Brandenburg \& Subramanian \cite{BrandSubr04})
\begin{equation}
{\rm St}^{\rm (rel)} = k_{f} u_{\rm rms} \tau_{\rm rel}\;,
\end{equation}
where $k_{f}$ is the wavenumber of the energy carrying scale and
$u_{\rm rms}$ the average rms-velocity in the convectively unstable
region. For convection calculations $k_{\rm f}$ corresponds
essentially to the largest possible scale permitted by the box
dimensions (see K\"apyl\"a et al. \cite{Kaepylaeea2004b})
\vfill

\subsection{Determination of the timescales}
\label{subsec:times}
The disadvantage of the MTA-approach is that it can only be applied to
the case where rotation and the Reynolds stresses are statistically
nonzero. In order to circumvent this problem, we have devised an
independent way to estimate St which works also if rotation is not
present and without any recourse to the mean-field theory. The method
essentially consists of the calculation of the correlation and
turnover times separately and directly from the flow.

The correlation time is estimated from the autocorrelation of velocity
\begin{eqnarray}
  C[u_{i}(\vec{x},t_0),u_{i}(\vec{x},t)] = \frac{ u_{i}(\vec{x},t_0)
    u_{i}(\vec{x},t)}{\sqrt{u_{i}^2(\vec{x},t_0)
      u_{i}^2(\vec{x},t)}}\;, \label{equ:corr}
\end{eqnarray}
where $\vec{x}$ is the position vector and $t_0$ and $t$ denote the
times from which the snapshots were taken. The correlation time
$\tau_{\rm c}$ is defined as the time after which the correlation
drops below a threshold value. In the present study the threshold is
set to 0.5. We use the vertical velocity component to determine the
correlation time.

The turnover time, $t_{\rm to}$, is estimated from the trajectories of
test particles which are advected by the flow. The turnover time can
be defined as the time which passes between two consecutive crossovers
of some fixed reference level into the same direction. However, this
definition implicitly assumes that the vertical scale of convective
motions is of the order of the depth of the convectively unstable
layer. Thus, turnovers happening far away in comparison to the scale
of convection are not registered at all. Another possibility is to
define the turnover time as the time which elapses between two
consecutive changes of direction (into the same direction). This
latter definition registeres all turnovers and we shall present
results using that in the remainder of the paper. Differences between
the results obtained with the two definitions are discussed further in
K\"apyl\"a et al.  (\cite{Kaepylaeea2004b}).

Once the correlation and turnover times have been determmined, the
Strouhal number is simply their ratio
\begin{equation}
{\rm St} = \frac{\tau_{\rm c}}{t_{\rm to}}\;.
\end{equation}

\begin{figure}
\resizebox{\hsize}{!}
{\includegraphics{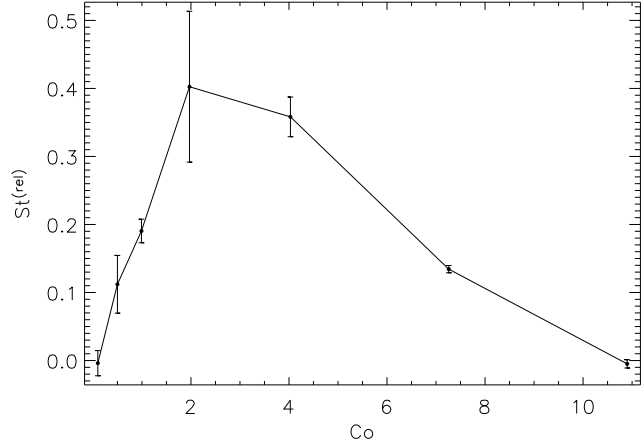}}
\caption{The Strouhal number from MTA.}
\label{fig:strel}
\end{figure}

\begin{figure*}[t]
\resizebox{\hsize}{!}
{\includegraphics{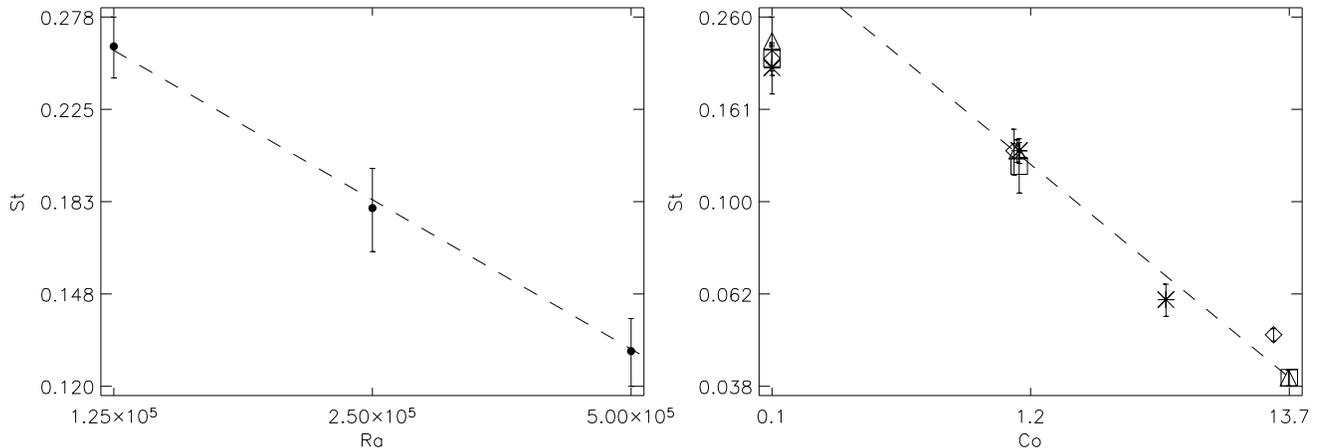}}
\vspace{-6.5cm}
\caption{The Strouhal number as function of the Rayleigh (left) and
  Coriolis (right) numbers. The dashed lines give the power laws ${\rm
    St} \propto {\rm Ra}^{-0.49}$ and ${\rm
    St} \propto {\rm Co}^{-0.45}$}
\label{fig:st}
\end{figure*}

\section{Results}
\label{sec:results}
The minimal tau-approximation can be applied to the Reynolds stresses
only if there is appreciable rotation, and away from the poles where
the stresses vanish due to symmetry in the present geometry
(e.g. K\"apyl\"a et al. \cite{Kaepylaeea2004a}). A typical result is
shown in Figure \ref{fig:strel}. The analysed calculations are those
made at a latitude 30 degrees south from K\"apyl\"a et
al. (\cite{Kaepylaeea2004a}). The stress is taken from the middle of
the convectively unstable layer which generally quite well describes
the situation in the whole box. The errorbars denote the modified mean
error of the stress (see Eq.~(30) of K\"apyl\"a et
al. \cite{Kaepylaeea2004a}). As function of the Coriolis number the
Strouhal number is essentially determined by the value of $\tau_{\rm
  rel}$, which varies in a very similar manner as the vertical
$\Lambda$-effect. The reason for this behaviour is that $u_{\rm rms}$
and the diagonal Reynolds stresses $Q_{\rm yy}$ and $Q_{\rm zz}$ vary
only little as function of Co leading to the fact that the functional
form for the relaxation time and the $\Lambda$-effect is essentially
the same, $Q_{yz}/\Omega$. The values of St$^{\rm (rel)}$ vary from
about 0.4 for Co = 2 to essentially zero for very slow and very rapid
rotation.

Whereas the applicability of the MTA-approach is limited to cases with
rotation, the Strouhal number should have a finite value also without
rotation. Thus, we set out to extract the correlation and turnover
times separately from the flow as described in
Sect.~(\ref{subsec:times}). We find that the the correlation time
decreases consistent with $\tau_{\rm c} \propto {\rm Ra}^{-0.45}$ as
function of the Rayleigh number. The turnover time, however, changes
only marginally. Thus the Strouhal number follows approximately a
power law ${\rm St} \propto {\rm Ra}^{-0.49}$ (see left panel of
Figure \ref{fig:st}). Although the parameter range in the present
study is quite limited, the result is still promising in the sense
that if the same trend carries over to stellar parameters, the
Strouhal number may be much smaller there.

A similar trend is found when rotation is increased. The correlation
time decreases rapidly as function of the Coriolis number whereas the
changes in the turnover time are only minor. However, one aspect not
to be overlooked here is the trend seen in the turnover time. The
simple estimate, the depth of the convectively unstable region divided
by the average velocity, increases as function of rotation due to the
fact that overall velocities tend to diminish as rotation becomes more
rapid. However, the turnover time calculated from the test particle
trajectories shows an opposite trend due to the fact that the spatial
scale of convection is reduced even more than the overall
velocities. Thus we find that for moderate and rapid rotation, the
Strouhal number approximately follows a power law ${\rm St} \propto
{\rm Ra}^{-0.45}$ (see right panel of Figure \ref{fig:st}). The
explanation is most probably the strong Coriolis forces which tend to
disrupt any coherent flow structures.

\section{Conclusions}
\label{sec:conclusions}
We estimate the Strouhal number from numerical models of convection
with two independent methods. Firstly, we apply the minimal
tau-approximation of the equation of the Reynolds stress. Secondly, we
calculate the correlation and turnover times directly from the flow
without any recourse to a mean-field theory.

We find that the Strouhal number from the MTA reaches values of
maximally $\approx 0.5$. St has a maximum for intermediate rotation,
where the Reynolds stress itself also peaks. The Strouhal number
follows closely the same trend as the vertical $\Lambda$-effect (see
e.g. K\"apyl\"a et al. \cite{Kaepylaeea2004a}) as function of rotation
due to the similar functional form.

The correlation time is seen to decrease consistent with power law
$\tau_{\rm c} \propto {\rm Ra}^{-0.45}$ as function of the Rayleigh
number and approximately with $\tau_{\rm c} \propto {\rm Co}^{-0.45}$
for moderate and rapid rotation. These results indicate that although
the values of St in this study are generally of the order of 0.1 to 1,
the Strouhal number in stellar convection zones, where Ra is most
certainly, and Co at least probably, much larger than in the present
study, may be significantly smaller.

\acknowledgements{PJK acknowledges the financial support from the
  Finnish graduate school for astronomy and space physics. PJK
  acknowledges the hospitality of NORDITA during his visit. MJK
  acknowledges the hospitality of LAOMP, Toulouse and the Kiepenheuer-
  Institut, Freiburg during her visits, and the Academy of Finland
  project 203366.}

\end{document}